\documentclass[aps,prb,twocolumn,superscriptaddress,showpacs,showkeys,amsmath]{revtex4-1}

\usepackage{amssymb,amsmath,graphicx}

\begin{document}





\title{Spatial distribution of dynamically polarized nuclear spins in electron spin domains in the $\nu = 2/3$ fractional quantum Hall state studied by nuclear electric resonance}



\author{S. Watanabe}
\email{wshinji@se.kanazawa-u.ac.jp}
\affiliation{ERATO Nuclear Spin Electronics Project, 468-15 Aoba, Aramaki-Aza, Aoba-ku, Sendai 980-8578, Japan}
\affiliation{Graduate School of Science, Department of Physics, Tohoku University, 6-3 Aoba, Aramaki-Aza, Aoba-ku, Sendai 980-8578, Japan}
\author{G. Igarashi}
\affiliation{Graduate School of Science, Department of Physics, Tohoku University, 6-3 Aoba, Aramaki-Aza, Aoba-ku, Sendai 980-8578, Japan}
\author{N. Kumada}
\affiliation{NTT Basic Research Laboratories, NTT Corporation, 3-1 Morinosato-Wakamiya, Atsugi 243-0198, Japan}
\author{Y. Hirayama}
\email{hirayama@m.tohoku.ac.jp}
\affiliation{ERATO Nuclear Spin Electronics Project, 468-15 Aoba, Aramaki-Aza, Aoba-ku, Sendai 980-8578, Japan}
\affiliation{Graduate School of Science, Department of Physics, Tohoku University, 6-3 Aoba, Aramaki-Aza, Aoba-ku, Sendai 980-8578, Japan}
\date{\today}

\begin{abstract}
Nuclear electric resonance (NER) is based on nuclear magnetic resonance mediated by spatial oscillations of electron spin domains excited by a radio frequency (RF) electric field, and it allows us to investigate the spatial distribution of the nuclear spin polarization around domain walls (DWs). Here, NER measurements were made of the dynamic nuclear spin polarization (DNP) at the spin phase transition of the fractional quantum Hall state at a Landau level filling factor of $\nu=2/3$. From the RF pulse power and pulse duration dependence of the NER spectrum, we show that the DNP occurs only within $\sim 100$ nm around DWs, and that it does not occur in DWs. We also show that DWs are pinned by the hyperfine field from polarized nuclear spins.
\end{abstract}





\maketitle

\section{\label{sec1}Introduction}

In a GaAs two-dimensional electron system (2DES), nuclear spins can be electrically polarized and detected through the contact hyperfine interaction. Dynamic nuclear spin polarization (DNP) in quantum Hall (QH) regime was first demonstrated by Wald \textit{et al.} \cite{wald1994local}. Flip-flop scattering caused by applying a DC bias between spin-resolved QH edge channels generates DNP that in turn modulates the effective Zeeman field and is detected by transport measurement. DNP in spin-resolved edge channels has also been reported in the fractional QH regime \cite{machida2002spin}. In a bulk 2DES, Kronm\"{u}ller \textit{et al.} found that DNP is generated when a large source-drain current is applied in the fractional QH state at the Landau level filling factor $\nu = 2/3$ \cite{kronmuller1998new,kronmuller1999new}, and DNP appears as a significant enhancement in the longitudinal resistance $R_{\textrm{xx}}$. Subsequent experiments\cite{smet2001ising,smet2002gate,kraus2002from,hashimoto2004nuclear,stern2004nmr} revealed that DNP is related to the formation of domain structures at the spin phase transition (SPT) at $\nu = 2/3$. DNP has also been observed in the breakdown regime of the quantum Hall effect\cite{kawamura2007electrical,kawamura2010spatial}, the submicron-scale region\cite{yusa2005controlled,kou2010dynamic}, and in a two subband system\cite{zhang2007nmr,guo2010probing}.

Since the nuclear spin polarization at the $\nu =2/3$ SPT can be sensitively detected through $R_{\textrm{xx}}$, it has been exploited for resistively-detected nuclear spin relaxation \cite{smet2002gate,hashimoto2002electrically,kumada2005spin,kumada2006low} and nuclear magnetic resonance (NMR)\cite{stern2004nmr,kumada2007nmr} measurements to study QH physics. However, the mechanism of DNP at $\nu = 2/3$ SPT and the spatial distribution of nuclear spin polarizations are still unclear\cite{kraus2002from,lok2004time,stern2004nmr}; it remains to be seen how nuclear spin polarization is related to microscopic domain structures at $\nu = 2/3$ SPT.

Recently, Kumada \textit{et al.} demonstrated nuclear spin resonance induced by a radio frequency (RF) electric field (nuclear electric resonance: NER) \cite{kumada2008electric} instead of an RF magnetic field. Applying an RF electric field to a gate at $\nu = 2/3$ causes spatial oscillations of electron spin domains, which generate an effective RF magnetic field locally around domain walls (DWs). Since the spatial distribution of the effective RF magnetic field reflects the structure of the domain wall (DW), NER measurements are useful for investigating the structures of electron spin domains and the spatial distribution of the nuclear spin polarizations.

In this study, we performed detailed NER measurements to reveal how the nuclear spin polarization is related to microscopic domain structures at $\nu = 2/3$. We showed the NER characteristics in terms of a large set of parameters, including the frequency and power of the RF electric field, RF pulse duration, and DC bias that is applied during the application of the RF electric field. The RF pulse power dependence of the NER spectra for the fundamental and second harmonic frequencies provides information on the spatial distribution of nuclear spin polarizations close to DWs. From these measurements, we concluded that DNP does not occur inside DWs. The spatial distribution of nuclear spin polarizations away from DWs can be determined from the dependence of the NER spectrum on the RF pulse duration. We show that DNP occurs only within $\sim 100$ nm around DWs. Moreover, we find that the DC bias power dependence of the NER spectrum is useful for investigating the DW motion under DNP. With it, we demonstrated pinning of DWs induced by the hyperfine field.

This paper is organized as follows. Section\,\ref{sec2} explains DNP at the $\nu=2/3$ spin phase transition. Section\,\ref{sec3} describes the device structure and the method of obtaining the NER signals. Section\,\ref{sec5} discusses the dependence of the NER spectrum on the frequency and the power of the RF electric field and on the RF pulse duration. Section\,\ref{sec7} presents the DC bias dependence of the NER spectrum, demonstrating the pinning effect of DWs. Section\,\ref{sec9} summarizes this paper.

\begin{figure}
\centering
\includegraphics{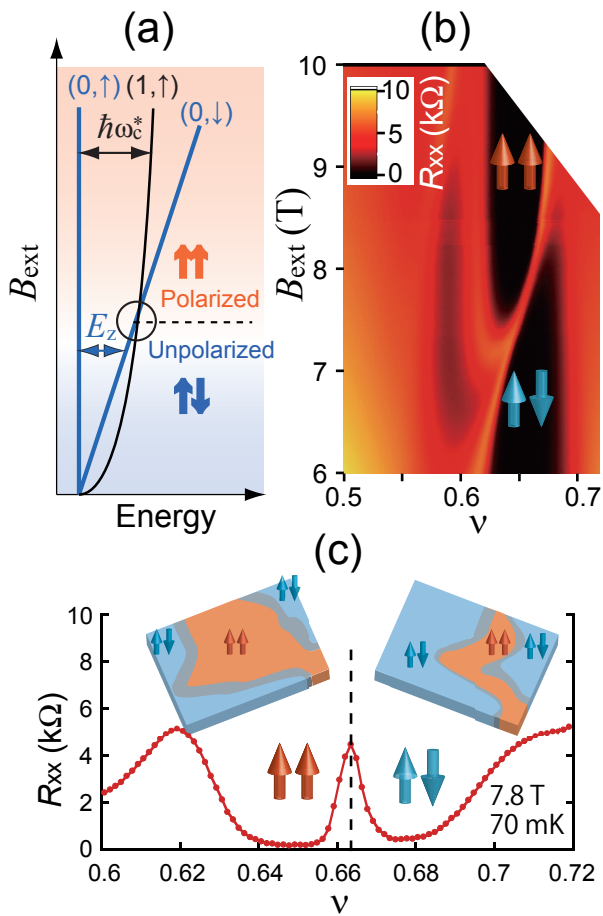}
\caption{(color online). (a) Schematic energy diagram of CF Landau levels as a function of the perpendicular magnetic field $B_{\textrm{ext}}$. (b) $R_{\textrm{xx}}$ as a function of $B_{\textrm{ext}}$ and $\nu$. (c)  $R_{\textrm{xx}}$ as a function of $\nu$ at 7.8 T. Spin-polarized and unpolarized states are formed in the minima at the lower and higher filling factor side of the SPT, respectively. The insets are schematic illustrations of the domain structures for $\nu$ values slightly higher and slightly lower than that for the SPT.
}
\label{fig1}
\end{figure}

\section{\label{sec2}DNP at the $\nu = 2/3$ spin phase transition}

Since this NER study is based on the formation of domain structures and DNP at $\nu = 2/3$ SPT, we briefly explain them before discussing our experiments. According to the composite fermion (CF) model \cite{jain1989composite}, a fractional QH state is regarded as an integer QH state of a CF, in which two flux quanta are attached to each electron. Figure \ref{fig1}(a) shows a schematic illustration of the energy diagram of CFs as a function of the magnetic field $B_{\textrm{ext}}$ applied perpendicular to a 2DES. Each level is characterized by the Landau level index of the CFs (N = 0, 1) and spin ($\uparrow$ or $\downarrow$). Two CF levels are filled at $\nu = 2/3$, which corresponds to $\nu$ = 2 of CFs. The phase transition between the spin-polarized and -unpolarized states occurs due to the competition between the Zeeman gap $E_{\textrm{Z}} (\propto B_{\textrm{ext}})$ and the CF cyclotron gap $\hbar \omega_{\textrm{c}} ^{\ast} (\propto \sqrt{B_{\textrm{ext}}})$\cite{einsenstein1990evidence,kukushkin1999spin}. A domain structure of these two different spin states is formed at the SPT.

Figure \ref{fig1}(b) shows $R_{\textrm{xx}}$ as a function of $B_{\textrm{ext}}$ and $\nu$. Around $\nu = 2/3$, the SPT appears as a resistance peak between the spin-unpolarized phase for smaller $B_{\textrm{ext}}$ and the spin-polarized phase for larger $B_{\textrm{ext}}$. In a certain range of fixed $B_{\textrm{ext}}$, the SPT occurs when $\nu$ changes [Fig.\,\ref{fig1}(c)]. This indicates that the domain structure can be changed by changing the gate bias: as the electron density $n$ and thus $\nu$ are increased around the SPT, the polarized domains shrink, whereas the unpolarized domains expand.

DNP occurs when a large current is applied at the $\nu = 2/3$ SPT, and it can be detected as an enhancement of $R_{\textrm{xx}}$ \cite{kronmuller1999new}. Although the detailed mechanisms of the DNP and the $R_{\textrm{xx}}$ enhancement remain unclear, they can be roughly explained as follows. When a current flows across the DWs, electron spins flip-flop scatter nuclear spins via the hyperfine interaction, leading to DNP. In this case, the spatial distribution of the current flow reflects the domain structure, and the resultant nuclear spin polarizations are spatially inhomogeneous \cite{stern2004nmr}. The inhomogeneous nuclear spin polarization acts as an inhomogeneous Zeeman field for electron spins, and this results in the enhancement of $R_{\textrm{xx}}$.

\section{\label{sec3}Sample and methods}

\begin{figure*}
\centering
\includegraphics{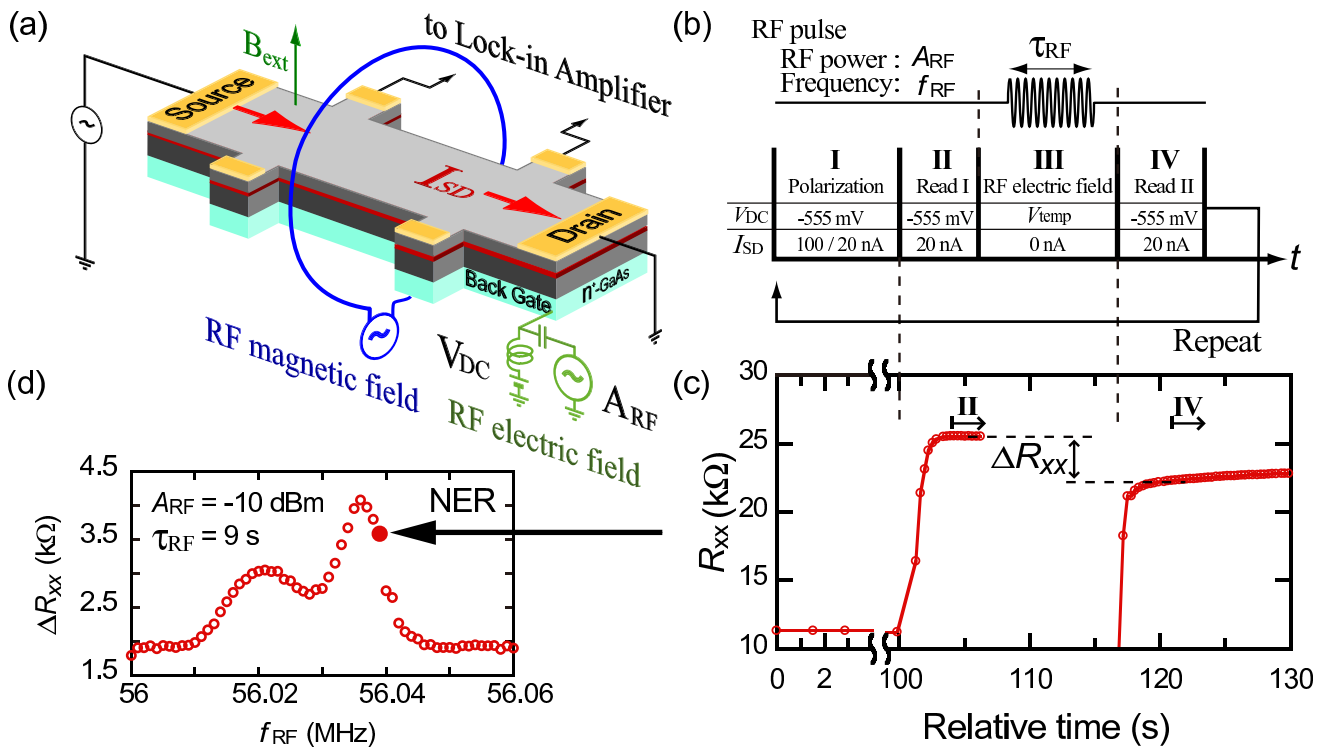}
\caption{(color online). (a) Schematic illustration of the experimental setup. (b) Sequence of the NER measurement. The table shows the values of $V_{\textrm{DC}}$ and $I_{\textrm{SD}}$ in each step. (c) An example of the $\Delta R_{\textrm{xx}}$ measurement (the parameters used in the measurement were $I_{\textrm{SD}} = 100$ nA for 100 s in step I, $A_{\textrm{RF}} = -10$ dBm, and $\tau_{\textrm{RF}} = 9$ s). $\Delta R_{\textrm{xx}}$ was determined by subtracting of the average of $R_{\textrm{xx}}$ over 2 s after the RF application (step IV) from that before (step II). $I_{\textrm{SD}}$ was suddenly changed before steps II and IV, and we waited for 3 s to obtain a correct value of $R_{\textrm{xx}}$ because of the slow time constant of the lock-in amplifier. The difference in $R_{\textrm{xx}}$ between steps I and II is due to the current dependent $R_{\textrm{xx}}$ at the $\nu = 2/3$ SPT after the DNP\cite{kraus2002from}. By repeating these steps while changing $f_{\textrm{RF}}$ (from $56.00$ to $56.06$ MHz in 1 kHz steps), we obtained the NER spectrum (d). The solid circle indicates $\Delta R_{\textrm{xx}}$ data obtained from Fig.\,(c).
}
\label{fig2}
\end{figure*}

The Hall-bar sample used in this study (50-$\mu$m-wide, 180-$\mu$m-long between the voltage leads) consists of a 20-nm-wide GaAs/AlGaAs quantum well. The electron density was controlled by using the $n^{+}$-GaAs substrate acting as a back gate. The low-temperature electron mobility is 1.8$\, \times \,$10$^{6}$cm$^{2}$/Vs at an electron density of 1.6$\, \times \,$10$^{15}$m$^{-2}$. 

$R_{\textrm{xx}}$ was measured using a standard low-frequency lock-in technique. We fixed the temperature and $B_{\textrm{ext}}$ at 150 mK and $7.685$ T, respectively. In the NER measurement, the RF electric field was superimposed on the DC bias $V_{\textrm{DC}}$ applied to the back gate [Fig.\,\ref{fig2}(a)]. The NER measurement consists of four steps [Fig.\,\ref{fig2}(b)]. In the first step (step I: Polarization), DNP was generated by setting the electronic system to the $\nu =2/3$ SPT ($V_{\textrm{DC}} = -555$ mV) and applying a large source-drain current $I_{\textrm{SD}}= 100$ or 20 nA for 100 s. In the second step (step II: Read I), $I_{\textrm{SD}}$ was set to 20 nA and $R_{\textrm{xx}}$ was measured for 2 s. In the third step (step III: RF electric field), $I_{\textrm{SD}}$ was set to zero ($V _{\textrm{temp}}= -555$ mV unless otherwise noted) and then an RF electric field with a frequency $f_{\textrm{RF}}$ and power $A_{\textrm{RF}}$ was applied to the back gate for a time duration $\tau_{\textrm{RF}}$. If $f_{\textrm{RF}}$ corresponds to the resonance frequency of the nuclear spins, they are depolarized. Note that, for the DC bias dependence of the NER spectrum described in Sec.\,\ref{sec7}, the system was tuned to a temporal filling factor by applying $V_{\textrm{temp}}$ during the RF application. In the fourth step (step IV: Read II), $R_{\textrm{xx}}$ at $V_{\textrm{DC}} = -555$ mV was measured for 2 s with $I_{\textrm{SD}} = 20$ nA. The size of the resistance drop $\Delta R_{\textrm{xx}}$, which is a measure of the nuclear spin depolarization induced by the RF application, was determined by subtracting the 2-s average of $R_{\textrm{xx}}$ in step IV from that in step II [Fig.\,\ref{fig2}(c)]. We obtained the NER spectrum [Fig.\,\ref{fig2}(d)] by plotting $\Delta R_{\textrm{xx}}$ as a function of $f_{\textrm{RF}}$. We also measured NMR spectra by using an NMR coil [Fig.\,\ref{fig2}(a)] as a reference. Note that we focused on spectra for $^{75}$As because its resonance intensity is the largest due to its natural abundance.

\section{\label{sec5}Spatial distribution of nuclear spin polarizations}

\subsection{\label{sec4-0}$A_{\textrm{RF}}$ dependence of the NER spectrum}

\begin{figure}[!t]
\centering
\includegraphics{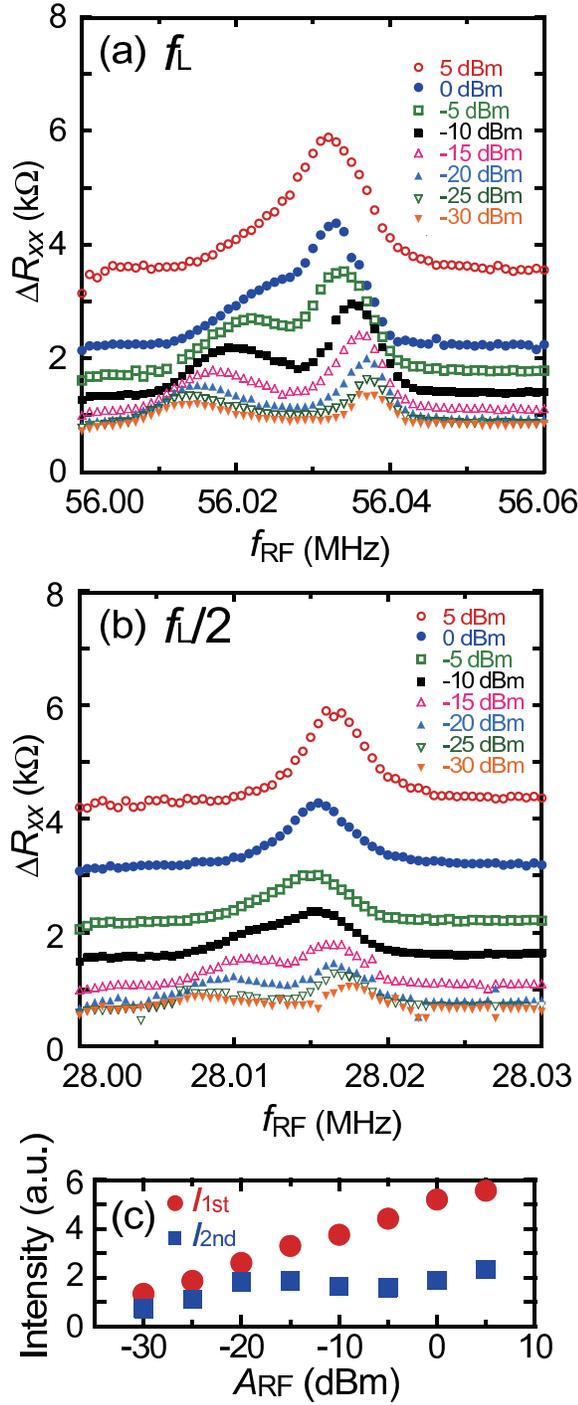}
\caption{(color online). NER spectra around $f_{\textrm{L}}$ (a) and $f_{\textrm{L}}/2$ (b) for several values of $A_{\textrm{RF}}$ between $-30$ to 5 dBm. The parameters were $I_{\textrm{SD}}$ = 100 nA for 100 s in step I and $\tau_{\textrm{RF}}$ = 5 s. (c) NER intensities around $f_{\textrm{RF}} = f_{\textrm{L}}$ ($I_{\textrm{1st}}$) and $f_{\textrm{L}}/2$ ($I_{\textrm{2nd}}$) as a function of $A_{\textrm{RF}}$. 
}
\label{fig3}
\end{figure}

In NER, applying an RF electric field to a gate causes spatial oscillations of the electron spin domains, which in turn generate a local oscillating hyperfine field around the DWs. The in-plane component of the oscillating hyperfine field acts as an effective RF magnetic field for the nuclear spins, while the out-of-component shifts the resonance frequency (Knight shift). Since the spatial extent of the DW oscillations can be changed by varying $A_{\textrm{RF}}$, the $A_{\textrm{RF}}$ dependence of the NER spectrum provides information on the spatial distribution of the nuclear spin polarizations around the DWs and on the structure of the DWs. Figure \ref{fig3}(a) shows NER spectra around the Larmor frequency of $^{75}$As, $f_{\textrm{L}}$ ( = 56.037 MHz  at $B_{\textrm{ext}}$ = 7.685 T) for several values of $A_{\textrm{RF}}$. For $A_{\textrm{RF}}$ = 5 dBm, a broad peak appears at 56.032 MHz. As $A_{\textrm{RF}}$ decreases, the peak splits into two peaks and their separation becomes large. The intensity $I_{\textrm{1st}}$ of the NER spectra \footnote{The NER signal intensity is obtained by integrating $\Delta R_{\textrm{xx}}$ after subtracting the background, which arises from heating and nuclear spin-lattice relaxation in the interval between step II and IV.} increases with $A_{\textrm{RF}}$[Fig.\,\ref{fig3}(c)]. Figure \ref{fig3}(b) shows NER spectra around $f_{L}/2$ for various $A_{\textrm{RF}}$. The intensity $I_{\textrm{2nd}}$ of the spectra around $f_{\textrm{L}}/2$ is comparable to $I_{\textrm{1st}}$, indicating that the effective RF magnetic field includes a large second-harmonic component.

\subsection{\label{sec4}Model}

\begin{figure}
\centering
\includegraphics{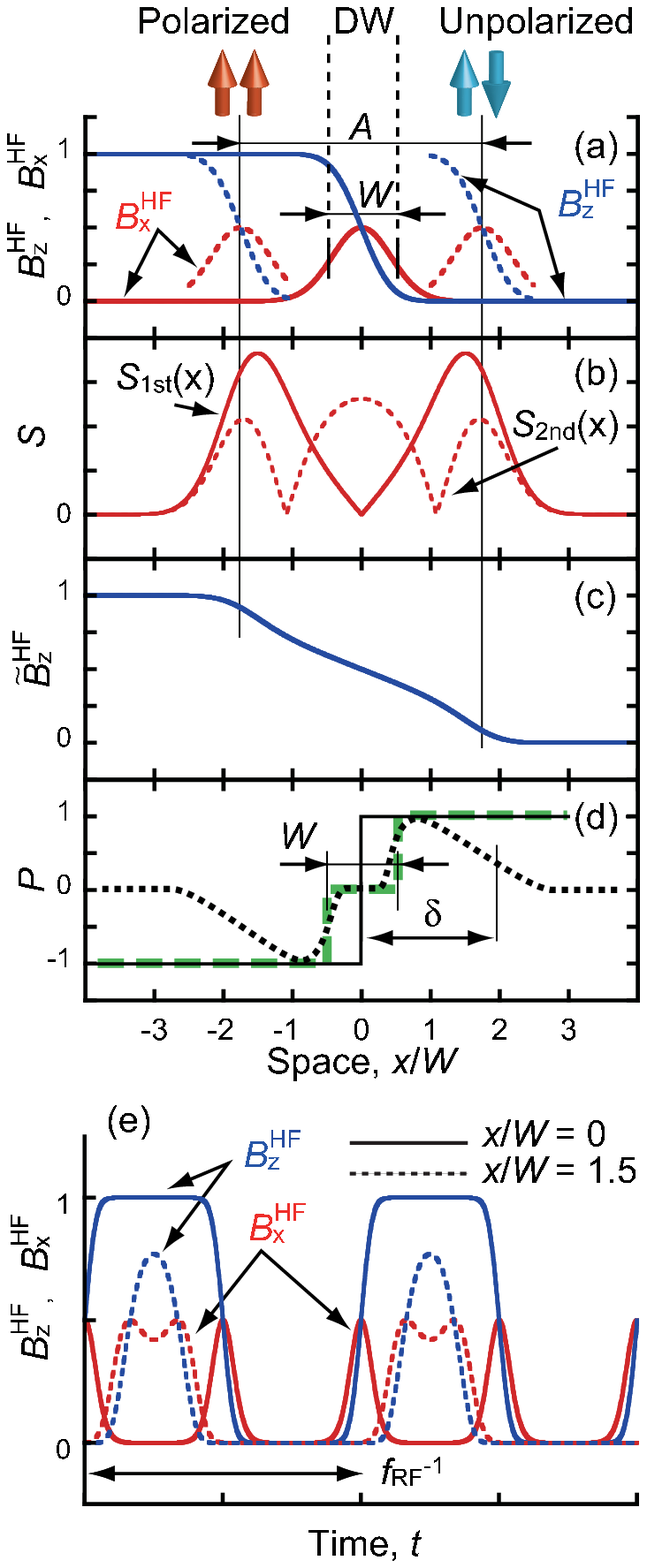}
\caption{(color online). (a) In-plane and out-plane components of the hyperfine field, $B_{\textrm{x}}^{\textrm{HF}}$ and $B_{\textrm{z}}^{\textrm{HF}}$ respectively, as a function of $x/W$. Dashed lines indicate $B_{\textrm{x}}^{\textrm{HF}}$ and $B_{\textrm{z}}^{\textrm{HF}}$ at moments in time $t = \pm 1/(4f_{\textrm{RF}})$. (b) Spectral density of $B_{\textrm{x}}^{\textrm{HF}}$  at the fundamental [$S_{\textrm{1st}}(x)$: solid line] and second-harmonic [$S_{\textrm{2nd}}(x)$: dashed line] components as a function of $x/W$. (c) Time average of $B_{\textrm{z}}^{\textrm{HF}}(t)$ over one period as a function of $x/W$. (d) Spatial distribution of nuclear spin polarizations $P(x)$ used for the fitting. Details are described in the main text. (e) $B_{\textrm{z}}^{\textrm{HF}}$ and $B_{\textrm{x}}^{\textrm{HF}}$ in the time domain at $x/W$ = 0 (solid lines) and $x/W$ = 1.5 (dashed lines).
}
\label{fig4}
\end{figure}

To analyze the NER spectra, we simulated spatially and temporally dependent hyperfine fields of DWs. We considered a DW at $x = x_{0}$ between spin-polarized ($x < x_{0}$) and spin-unpolarized ($x > x_{0}$) domains. The out-of-plane component of the hyperfine field changes between the values for the full ($B_{\textrm{z}}^{\textrm{HF}} = 1$) and null ($B_{\textrm{z}}^{\textrm{HF}} = 0$) spin polarizations across the DW. In contrast, the in-plane component of the hyperfine field $B_{\textrm{x}}^{\textrm{HF}}$ is finite only in the DW; it reaches a maximum at $x = x_{0}$. We assumed that the shape of $B_{\textrm{x}}^{\textrm{HF}}$ was a Gaussian function with a full width at half maximum $W$:
\begin{equation}
B_{\textrm{x}}^{\textrm{HF}}(x) \propto \exp \{ -4\ln 2 \frac{(x-x_{0})^{2}}{W^{2}} \}.
\label{Bx(x)}
\end{equation}
Note that $W$ represents the width of DW and it is theoretically estimated to be $4l_{\textrm{B}}$\cite{shibata2007phase} with $l_{\textrm{B}}$ being the magnetic length; $W \sim 40$ nm at $B_{\textrm{ext}} = 7.685$ T. $B_{\textrm{z}}^{\textrm{HF}}$ is given by
\begin{equation}
B_{\textrm{z}}^{\textrm{HF}}(x)  = \alpha \times \frac{1}{2}\sqrt{1-\left[ \frac{B_{\textrm{x}}^{\textrm{HF}}(x)}{B_{\textrm{x}}^{\textrm{HF}}(x_{0})}\right]^{2}} + \frac{1}{2},
\label{Bz(x)}
\end{equation}
where $\alpha = 1$ for $x \leq x_{0}$ and $\alpha = -1$ for $x_{0} < x$.
By applying a sinusoidal RF electric field with a frequency $f_{\textrm{RF}}$ and a power $A_{\textrm{RF}}$ to the back gate [Fig \ref{fig2}(a)], the DW spatially oscillates with an amplitude $A$ that is determined by $A_{\textrm{RF}}$; i.e., $x_{0}(t)$ = $\frac{A}{2}\sin 2\pi f_{\textrm{RF}} t$ [Fig.\,\ref{fig4}(a)]. The spatial oscillations of the DW induce temporal oscillations of $B_{\textrm{z}}^{\textrm{HF}}$ and $B_{\textrm{x}}^{\textrm{HF}}$ around the DW. Figure \ref{fig4}(e) shows examples of time domain signals of  $B_{\textrm{z}}^{\textrm{HF}}$ and $B_{\textrm{x}}^{\textrm{HF}}$ at $x/W = 0$ (solid lines) and $x/W = 1.5$ (broken lines) with $A/W = 3.5$. At $x/W = 0$, $B_{\textrm{x}}^{\textrm{HF}}(t)$ has sharp peaks spaced at equal intervals. At $x/W = 1.5$, the dwell time in the spin-unpolarized domain is long and the interval between the $B_{\textrm{x}}^{\textrm{HF}}(t)$ peaks changes. $B_{\textrm{z}}^{\textrm{HF}}(t)$ is a square wave function with a position dependent duty ratio.

We devised a function for fitting the NER spectra. We assumed that the resonance strength of NER is determined by the product of the spectral density $S(x)$ of $B_{\textrm{x}}^{\textrm{HF}}(t)$ at the resonance frequency and the degree of nuclear spin polarization $\vert P(x) \vert$ as a function of $x$. The resonance frequency reflects the Knight shift $K(x)$, which is proportional to the time average of $B_{\textrm{z}}^{\textrm{HF}}(t)$ over one period ($\tilde{B}_{\textrm{z}}^{\textrm{HF}}$); $K(x) = \tilde{B}_{\textrm{z}}^{\textrm{HF}} (x) \times K$, where $K=25$ kHz is the Knight shift for the fully polarized state \footnote{We measured the NMR spectrum at $\nu=1$ and obtained the Knight shift $K(\nu=1)$ at $\nu=1$. $K$ is determined to be $K=2/3 \times K(\nu=1)$}. The fitting function \footnote{In our sample, the NMR spectrum in the absence of the 2DES shows a quadrupole splitting of $\sim$ 3 kHz. However, Eq.\,(\ref{I(f)}) does not take into account quadrupole splitting because the intensities of the splitting peaks are very small compared with the intensity of the center peak.} is given by
\begin{equation}
\begin{split}
I (f_{\textrm{RF}}) &=
\int_{f_{\textrm{L}}}^{f_{\textrm{L}}+K(x)} dz_{\textrm{f}} \int dx
\\ &\sqrt{\frac{z_{\textrm{f}}-f_{\textrm{L}}}{f_{\textrm{L}} + K(x) -z_{\textrm{f}}}}
g(\Gamma; f_{\textrm{RF}}-z_{\textrm{f}}) \vert P(x) \vert S(x),
\label{I(f)}
\end{split}
\end{equation}
where $g(\Gamma; f_{\textrm{RF}} - z_{\textrm{f}})$ is a Gaussian function with width $\Gamma$, which mainly arises from nuclear dipolar broadening of $\sim 3$ kHz. The first integral represents the contribution of the spatially varying electron density due to a finite QW thickness\cite{kuzma1998ultraslow}. $S(x)$ is obtained by Fourier transformation of $B_{\textrm{x}}^{\textrm{HF}}(t)$. Figure \ref{fig4}(b) shows the spectral density at the fundamental [$S_{\textrm{1st}}(x)$] and second-harmonic [$S_{\textrm{2nd}}(x)$] components as a function of $x$. $S_{\textrm{1st}}(x)$ has two peaks at $\vert x \vert \sim A/2$, whereas $S_{\textrm{2nd}}(x)$ has three peaks at $x = 0$ and $\vert x \vert \sim A/2$. Note that the absence of $S_{\textrm{1st}}(x=0)$ and the large value of $S_{\textrm{2nd}}(x=0)$ come from the fact that the interval between the $B_{\textrm{x}}^{\textrm{HF}}(t)$ peaks is $(2f_{\textrm{RF}})^{-1}$ at $x=0$ [Fig.\,\ref{fig4}(e)]. As shown in Fig.\,\ref{fig4}(c), $\tilde{B}_{\textrm{z}}^{\textrm{HF}}$ changes almost linearly for $\vert x \vert < A/2$

\subsection{\label{sec4-1}Analysis}

\begin{figure}[!t]
\centering
\includegraphics{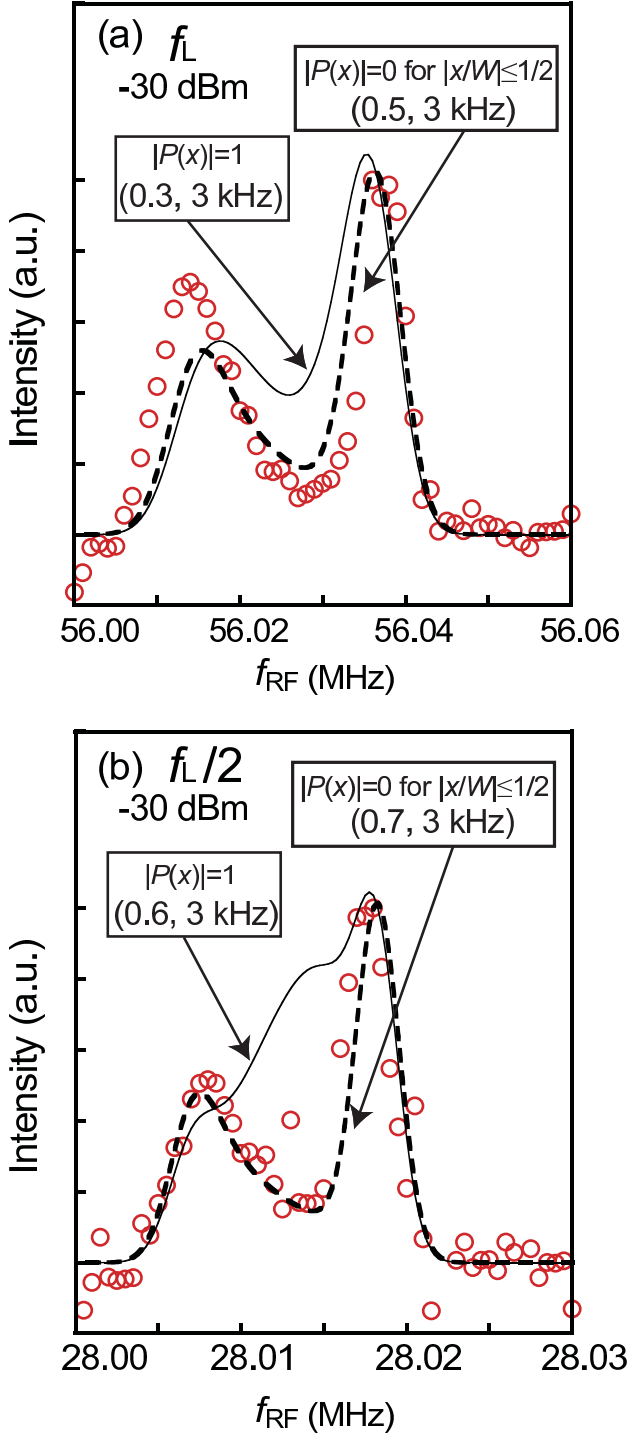}
\caption{(color online). Results of the fitting for the NER spectra around $f_{\textrm{L}}$ (a) and $f_{\textrm{L}}/2$ (b) at $A_{\textrm{RF}} = -30$ dBm. Solid lines are obtained under the assumption $\vert P(x) \vert =1$, while the dashed lines are for $\vert P(x) \vert =1$ for $\vert x/W \vert >1/2$ and $\vert P(x) \vert =0$ for $\vert x/W \vert \leq 1/2$. The fitting parameters ($A/W$, $\Gamma$) are indicated in the figures.
}
\label{fig5}
\end{figure}

\begin{figure}[!t]
\centering
\includegraphics{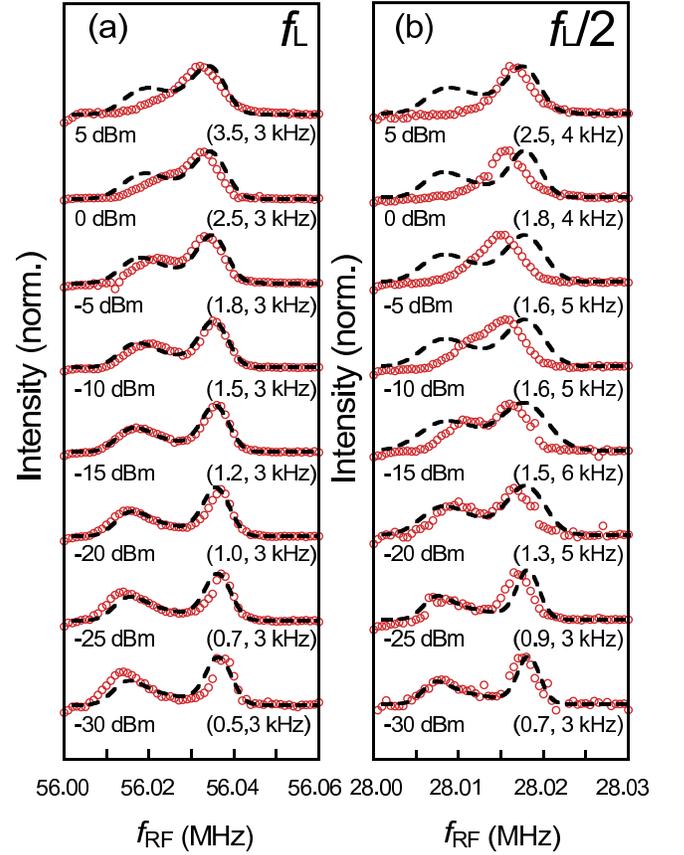}
\caption{(color online). Fitting results for the NER spectra around $f_{\textrm{L}}$ (a) and $f_{\textrm{L}}/2$ (b) for several values of $A_{\textrm{RF}}$. The NER spectra are normalized by the peak height for easy comparison. Dashed lines are for $\vert P(x) \vert =1$ for $\vert x/W \vert >1/2$ and $\vert P(x) \vert =0$ for $\vert x/W \vert \leq 1/2$. The fitting parameters ($A/W$, $\Gamma$) are indicated in the figures. 
}
\label{fig6}
\end{figure}

\begin{figure*}[!t]
\centering
\includegraphics{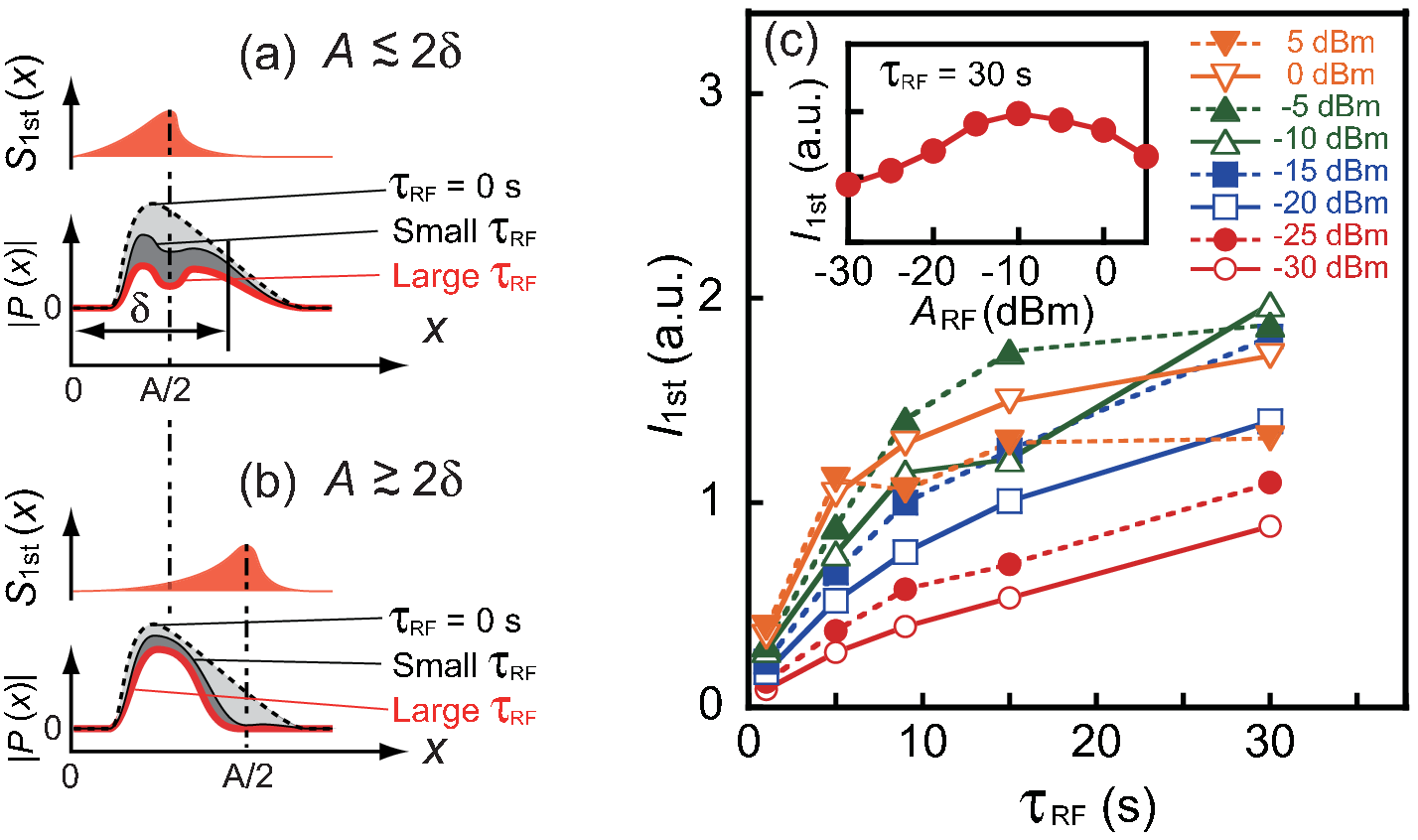}
\caption{(color online). Schematic illustrations of the time evolution of $\vert P(x) \vert$ for $A \lesssim 2\delta$ (a) and $A \gtrsim 2\delta$ (b). Dashed, thin, and bold lines indicate $P(x)$ for just after DNP ($\tau_{\textrm{RF}}$ = 0 s), small $\tau_{\textrm{RF}}$, and large $\tau_{\textrm{RF}}$, respectively. The shaded area corresponds to the size of $I_{\textrm{1st}}$. (c) The $\tau_{\textrm{RF}}$ dependence of $I_{\textrm{1st}}$ for several values of $A_{\textrm{RF}}$. All data were taken by using $I_{\textrm{SD}} = 100$ nA for 100 s in step I. The inset shows the dependence of $I_{\textrm{1st}}$ on $A_{\textrm{RF}}$ for $\tau_{\textrm{RF}}=30$ s.
}
\label{fig7}
\end{figure*}


We will discuss the spatial distribution of the nuclear spin polarizations around DWs on the basis of Eq.\,(\ref{I(f)}). Since the DNP is caused by flip-flop scattering upon electron transport across DWs, the nuclear spins are expected to be polarized around the DWs in the direction opposite to the electron spins at the Fermi level. Since the hyperfine field $B_{\textrm{N}}$ [$\propto P(x)/g,$ with $g$ the electronic $g$ factor of GaAs $\sim -0.44$\cite{weisbuch1977optical}] modulates $E_{\textrm{Z}}$ ($\propto B_{\textrm{N}}$), we can expect that $P(x) < 0$ and $P(x) > 0$ in spin-polarized ($x<0$) and -unpolarized ($x>0$) domains, respectively [Fig.\,\ref{fig4}(a)]. First, let us consider the case in which the nuclear spins are polarized only in the DWs; $\vert P(x) \vert = 1$ for $\vert x/W \vert < 1/2$ and $\vert P(x) \vert = 0$ for $\vert x/W \vert \geq 1/2$. In this case, since $S_{\textrm{1st}}(x) = 0$, and while $S_{\textrm{2nd}}(x)$ is at a maximum at $x=0$ [Fig.\,\ref{fig4}(b)], $I_{\textrm{1st}}$ is much smaller than $I_{\textrm{2nd}}$. This behavior is inconsistent with the experimental results [Fig.\,\ref{fig3}(c)]. Next, let us consider the case in which the nuclear spins are polarized in a wide region around the DWs. For simplicity, we will assume $\vert P(x) \vert = 1$ for all $x$ [solid line in Fig.\,\ref{fig4}(d)]. The solid lines in Figs.\,\ref{fig5}(a) and (b) are the fitting results of $\vert P(x) \vert = 1$ for the NER spectra around $f_{\textrm{RF}}=f_{\textrm{L}}$ and $f_{\textrm{L}}/2$, respectively, for a small RF power $A_{\textrm{RF}}=-30$ dBm; here, $\Gamma$ and $A/W$, by which $S_{\textrm{1st}}(x)$ is determined, are adjustable parameters. Around $f_{\textrm{RF}}=f_{\textrm{L}}$, the fitting reproduces the presence of the two resonance peaks, although it overestimates the intensity between them. The two peaks stem from the two peaks of $S_{\textrm{1st}}(x)$ and the different values of $\tilde{B}_{\textrm{z}}^{\textrm{HF}}$ at $x \sim \pm A/2$. Around $f_{\textrm{RF}}= f_{\textrm{L}}/2$, on the other hand, the fitting\footnote{The fittings around $f_{\textrm{RF}}=f_{\textrm{L}}/2$ are obtained by replacing $f_{\textrm{L}}$ and $K$ by $f_{\textrm{L}}/2$ and $K/2$, respectively, in Eq.\,(\ref{I(f)})} has an unwanted additional peak between the two. This is due to the $S_{\textrm{2nd}}(x)$ peak at $x=0$. To suppress the additional peak, we set the nuclear spin polarization inside the DW to zero; $\vert P(x) \vert = 0$ for $\vert x/W \vert \leq 1/2$ and $\vert P(x) \vert = 1$ for $\vert x/W \vert > 1/2$ [dashed line in Fig.\,\ref{fig4}(d)]. This greatly improves the fitting [dashed lines in Figs.\,\ref{fig5}(a) and (b)]. In particular, the frequencies and shapes of the two resonance peaks around $f_{\textrm{RF}} = f_{\textrm{L}}/2$ are well reproduced.

The fitting results indicate that the DNP does not occur inside the DWs. We suggest that this is due to a faster nuclear spin relaxation in the DWs. The in-plane component of the electron spins is finite in the DWs and it supports low-energy spin fluctuation\cite{Fal'ko1999Topological} at the resonance frequencies of nuclear spins. As a result, the nuclear spin-lattice relaxation rate becomes larger than the pumping rate of the DNP and the nuclear spins remain in thermal equilibrium.

The fitting results for the whole range of $A_{\textrm{RF}}$ provide further insight into $P(x)$, particularly away from DWs. The dashed lines in Figs.\,\ref{fig6}(a) and (b) indicate fittings assuming $\vert P(x) \vert =0$ for $\vert x/W \vert \leq 1/2$ and $\vert P(x) \vert =1$ for $\vert x/W \vert >1/2$. For smaller $A_{\textrm{RF}}$, these fittings reasonably well reproduce the experimental spectra around $f_{\textrm{RF}}=f_{\textrm{L}}$ and $f_{\textrm{L}}/2$. However, for larger $A_{\textrm{RF}}$, the NER spectra have only a single peak, whereas the fittings have two peaks.

We suggest that the discrepancy in the line shape of the NER spectra between the experiments and the simulations for large $A_{\textrm{RF}}$ mainly comes from electron heating of the 2DES. The application of the RF electric field increases the 2DES temperature, resulting in a decrease (increase) in the electron spin polarization of the spin-polarized (-unpolarized) domains. Since the separation between the two peaks of the NER spectra is larger for larger values of $\vert \tilde B_{\textrm{z}}^{\textrm{HF}}(x \sim A/2)$ - $\tilde B_{\textrm{z}}^{\textrm{HF}}(x \sim -A/2) \vert$, the electron heating reduces the separation between the two peaks. The NER spectrum ends up with a single broad peak when the electron heating is very large.

Information on $P(x)$ away from DWs can be derived from the NER spectra even when the electron heating is large for large $A_{\textrm{RF}}$. $\vert P(x) \vert$ away from DWs is expected to decay with a characteristic distance $\delta$ [broken line in Fig.\,\ref{fig4}(d)] that is determined by the competition between the nuclear spin diffusion rate and the nuclear spin-lattice relaxation rate $T_{1}^{-1}$. Here, $\delta$ can be experimentally obtained by analyzing the dependence of the NER spectrum on $\tau_{\textrm{RF}}$. As shown in the previous subsection, the strength of the resonance around $f_{\textrm{RF}}=f_{\textrm{L}}$ is determined by the product of $P(x)$ and $S_{\textrm{1st}}(x)$. In addition, $\tau_{\textrm{RF}}$ affects the NER spectrum. For $A \lesssim 2\delta$, $\vert P( x \sim A/2)\vert$ decreases with increasing $\tau_{\textrm{RF}}$ [Fig.\,\ref{fig7}(a)]. Thus, $I_{\textrm{1st}}$ increases almost linearly with $\tau_{\textrm{RF}}$. In contrast, for $A \gtrsim 2\delta$, $\vert P(x \sim A/2)\vert$ becomes almost zero at large $\tau_{\textrm{RF}}$ [Fig.\,\ref{fig7}(b)]. This absence of $\vert P(x \sim A/2)\vert$ suppresses the increase in $I_{\textrm{1st}}$. Since $A$ is determined by $A_{\textrm{RF}}$, the $\tau_{\textrm{RF}}$ dependence of $I_{\textrm{1st}}$ can be used to investigate $P(x)$ away from the DWs. Figure \ref{fig7}(c) shows the dependence of $I_{\textrm{1st}}$ on $\tau_{\textrm{RF}}$ from 1 to 30 s for several values of $A_{\textrm{RF}}$. For $A_{\textrm{RF}} = -30$ dBm, $I_{\textrm{1st}}$ increases almost linearly with $\tau_{\textrm{RF}}$. As $A_{\textrm{RF}}$ increases further, however, the dependence of $I_{\textrm{1st}}$ on $\tau_{\textrm{RF}}$ deviates from a linear one and $I_{\textrm{1st}}$ seems to approach an asymptotic value after $\tau_{\textrm{RF}}=30$ s. These asymptotic values decrease with increasing $A_{\textrm{RF}}$ [inset of Fig.\,\ref{fig7}(c)]. This behavior is very different from that for $\tau_{\textrm{RF}}=5$ s [Fig.\,\ref{fig3}(c)].

The deviation of $I_{\textrm{1st}}$ from being approximately linearly dependent on $\tau_{\textrm{RF}}$ starts around $A_{\textrm{RF}} \sim -5$ dBm, suggesting that $\delta$ exists in this region. More quantitatively, from the inset of Fig.\,\ref{fig7}(c), we can roughly estimate $\delta$ as the peak position of $I_{\textrm{1st}}$ ($A_{\textrm{RF}}=-10$ dBm). The fitting results shown in Fig.\,\ref{fig6}(a) ($A/W \sim A/$40 nm = 1.5 at $A_{\textrm{RF}}=-10$ dBm) give a $\delta$ of $A/2 \sim 30$ nm. We can compare this value with an estimate based on nuclear spin diffusion and $T_{1}$. For a sufficiently long period of time after DNP, the spatial distribution of nuclear spin polarizations will have spread out. In such a case, the diffusion radius would be equivalent to $\delta$, which can be roughly estimated to be $\sqrt{DT_{1}} \sim 30 - 100$ nm, where $D \sim 10$ nm$^{2}$/s is the diffusion constant for $^{75}$As nuclei in GaAs crystals\cite{paget1982optical}, and $T_{1}$ would be on the order of $100$ s\cite{barrett1995optically,hashimoto2002electrically} at 150 mK. This value is of the same order as our experimentally obtained value.

\section{\label{sec7}Pinning of the domain walls}

\begin{figure*}[!t]
\centering
\includegraphics{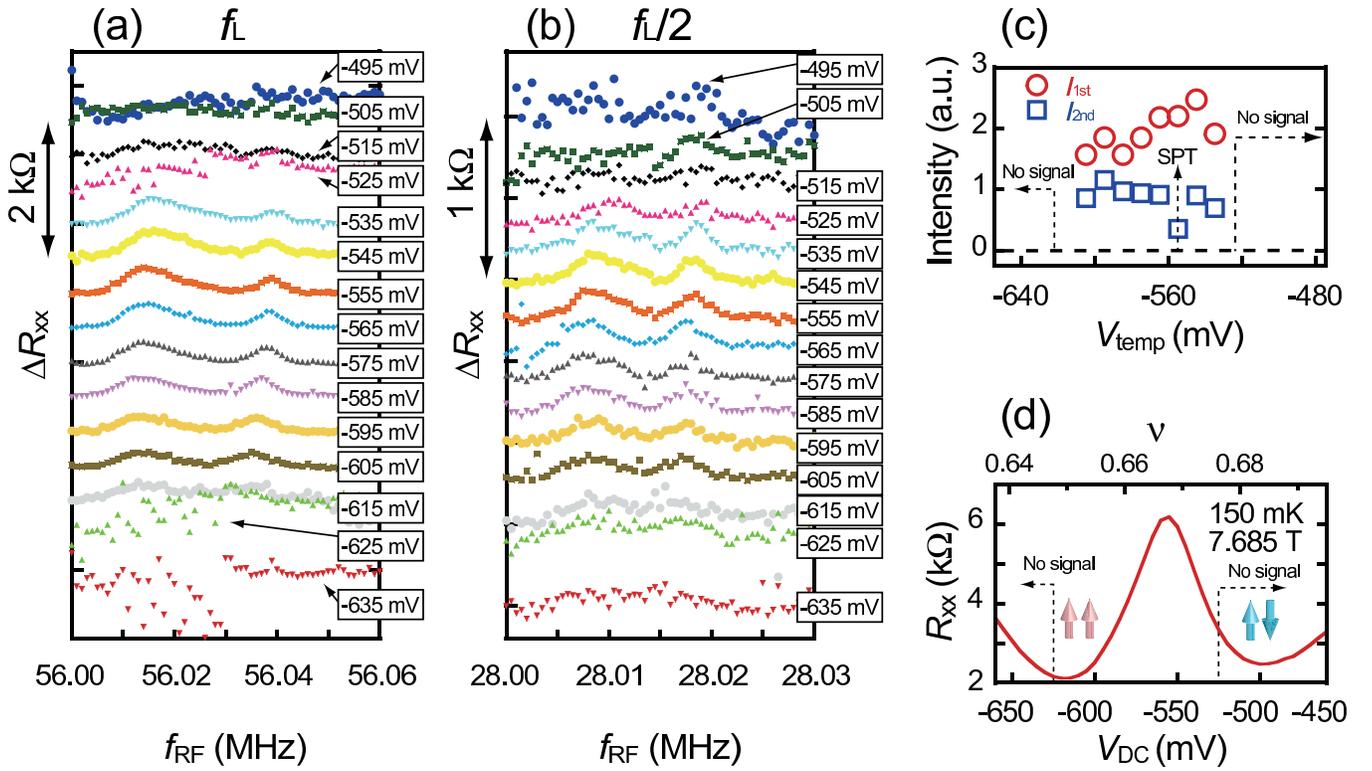}
\caption{(color online). NER spectra around $f_{\textrm{RF}}=f{\textrm{L}}$ (a) and $f_{\textrm{L}}/2$ (b) plotted for several values of $V_{\textrm{temp}}$. Each trace is vertically offset for clarity. To obtain these data, we used $I_{\textrm{SD}} = 20$ nA for 800 s at $\nu = 2/3$ before step I to saturate the DNP. The parameters are $I_{\textrm{SD}} = 20$ nA, $A_{\textrm{RF}} = -30$ dBm, and $\tau_{\textrm{RF}} = 5$ s. (c) NER intensities around $f_{\textrm{RF}}=f_{\textrm{L}}$ and $f_{\textrm{L}}/2$ obtained from (a) and (b). (d) $R_{\textrm{xx}}$ data at $B_{\textrm{ext}}$ = 7.685 T around $\nu = 2/3$.
}
\label{fig8}
\end{figure*}

Finally, let us investigate how the NER spectrum changes when $V_{\textrm{temp}}$ changes around the SPT at $V_{\textrm{DC}} = -555$ mV. Figures \ref{fig8}(a) and (b) show the spectra for several values of $V_{\textrm{temp}}$ around $f_{\textrm{RF}} = f_{\textrm{L}}$ and $f_{\textrm{L}}/2$, respectively. For $-615$ mV $\lesssim V_{\textrm{temp}} \lesssim -525$ mV, the shape and intensity of the spectra are almost independent of $V_{\textrm{temp}}$ [Fig. \ref{fig8}(c)]. However, as $V_{\textrm{temp}}$ changes further away from the SPT for $V_{\textrm{temp}} \lesssim -625$ mV and $V_{\textrm{temp}} \gtrsim -515$ mV, the NER signal suddenly disappears around $f_{\textrm{L}}$ and around $f_{\textrm{L}}/2$.

The disappearance of the NER signal for $V_{\textrm{temp}}$ away from the SPT is due to the disappearance of the domain structure. As shown in Fig.\,\ref{fig8}(d), $R_{\textrm{xx}}$ has minima at $V_{\textrm{DC}} = -625$ and $-500$ mV, at which the system is covered with the spin-polarized and -unpolarized phases, respectively. The disappearance of the NER signals demonstrates the essential role of the domain structure in NER.

For $-615$ mV $\lesssim V_{\textrm{temp}} \lesssim -525$ mV, the NER spectrum is naively expected to depend on $V_{\textrm{temp}}$ because the position of the DWs and thus $S(x)$ should be modulated by $V_{\textrm{temp}}$. The observed independence in turn suggests that DWs are pinned by polarized nuclear spins around the DWs. Since the spin-polarized (-unpolarized) domains are stabilized by larger (smaller) $B_{\textrm{N}}$, a large difference in $B_{\textrm{N}}$ between both sides of the DWs [Fig.\,\ref{fig4}(d)] fixes the position of the DWs. As a result, the domain structure is unchanged by a small variation of $V_{\textrm{temp}}$. Note that the pinning effect can be controlled electrically by changing the degree of the DNP.

We stress that pinning is ineffective under an RF electric field even when $A_{\textrm{RF}}$ is small. That is, under the RF electric field, small spatial oscillations of DWs gradually destroy the nuclear spin polarizations $near$ DWs even if there is some pinning. The decrease in the nuclear spin polarizations decreases the pinning effect. The pinning under the RF electric field eventually becomes small, and this means our simple model, in which the displacement of DWs is determined by $A_{\textrm{RF}}$, is valid.

\section{\label{sec9}summary}

We investigated the spatial distribution of dynamically polarized nuclear spins in electron spin domains at the $\nu = 2/3$ spin phase transition by using NER. The NER spectra for the fundamental and second harmonic frequencies depend on the RF pulse power $A_{\textrm{RF}}$ and the RF pulse duration. A model based on DW oscillations induced by the RF electric field explains the NER spectra. From a detailed analysis, we concluded that the DNP occurs only within $\sim 100$ nm around DWs, not inside them. The NER spectra are almost independent of the DC bias that is applied around $\nu = 2/3$ during application of the RF electric field. This suggests that the DWs are pinned by the hyperfine field produced by the polarized nuclear spins around the DWs. Our findings demonstrate that NER is useful for revealing how microscopic domain structures are related to the nuclear spin polarization around DWs and imply the possibility of electric manipulation of spatial domain structures and nuclear spin polarization at the nanometer scale.

\begin{acknowledgments}

The authors are grateful to S. Miyashita and T. Kobayashi for the epitaxial growth and sample processing, and K. Akiba for helpful discussions. We also thank the Germany-Japan Cooperative Program of JST.

\end{acknowledgments}



%

\end{document}